\documentclass[11pt]{article}

\usepackage{amsfonts}
\usepackage{amsmath}
\usepackage[dvips]{graphicx}
\usepackage{graphicx,indentfirst,amsmath}
\usepackage{caption}
\usepackage{subcaption}
\usepackage{graphics}
\usepackage{comment}
\usepackage{lscape}
\usepackage{rotating}
\usepackage{color}
\usepackage{bm}
\usepackage[round]{natbib}

\makeatletter

\newcommand{\Rmnum}[1]{\expandafter\@slowromancap\romannumeral #1@}
\makeatother

\def\<{\langle}
\def\>{\rangle}

\begin{document}

  \title{\bf Penalized composite likelihood for colored graphical Gaussian models}
  \author{Qiong Li 
  \thanks{To whom correspondence should be addressed. Email: liqiong29@mail.sysu.edu.cn} \ \ \
  Xiaoying Sun \textsuperscript{b}\ \ \ Nanwei Wang \textsuperscript{\dag}\\
 % \thanks{To whom correspondence should be addressed. Email: wangnanw@yorku.ca}
%   \ \ \  Xin Gao  \thanks{To whom correspondence should be addressed. Email: xingao@mathstat.yorku.ca}\\
   %\thanks{Supported in part by the Natural Sciences and Engineering Research Council of Canada (NSERC) individual discovery grants.}\\
    \textsuperscript{*} School of Mathematics (Zhuhai), Sun Yat-sen University, Zhuhai, \\519082, Guangdong, China\\
    \textsuperscript{b} Department of Mathematics and Statistics, York University, \\Toronto, M3J1P3, Canada \\
    \textsuperscript{\dag} Lunenfeld-Tanenbaum Research Institute, Mount Sinai Hospital,\\ Toronto, M5G1X5, Canada}
  \maketitle

\abstract
{This paper proposes a penalized composite likelihood method for model selection in colored graphical Gaussian models. The method provides a sparse and symmetry-constrained estimator of the precision matrix, and thus conducts model selection and precision matrix estimation simultaneously. In particular, the method uses penalty terms to constrain the elements of the precision matrix, which enables us to transform the model selection problem into a constrained optimization problem. Further, computer experiments are conducted to illustrate the performance of the proposed new methodology. It is shown that the proposed method performs well in both the selection of nonzero elements in the precision matrix and the identification of symmetry structures in graphical models.
The feasibility and potential clinical application of the proposed method are demonstrated on a microarray gene expression data set. }


\section{Introduction:}

In recent years, undirected graphical models \citep{lauritzen1996graphical} have been playing an important role in statistical inference,
which are widely employed to analyze and visualize conditional dependence relationships among variables. In a graphical model of a multivariate distribution, vertices represent random variables and edges encode conditional dependencies among the vertices. Precision matrix estimation and model selection in graphical Gaussian models is equivalent to estimating parameters and identifying zeros in the precision matrix.
The increasing availability of large data in different disciplines
makes graphical models an excellent tool to capture the conditional structure
between component variables. Graphical Gaussian models have been successfully applied in a number of fields such as genetic networks \citep{dobra2004sparse}, biological networks
\citep{newman2003structure} and financial networks \citep{fan2012vast}.

Colored graphical models are developed by adding symmetry restrictions to the precision matrix of graphical models \citep{Hoj08}. As constrainted graphical Gaussian models, colored graphical models can be represented by coloring the associated underlying graphs. The colored edges and vertices are associated with the restricted equal entries in the precision matrix. Adding symmetry restriction to the precision matrix reduces the number of parameters, and thus is useful when the number of variables greatly exceeds the number of observations. In \cite{Hoj08}, the authors derived a maximum likelihood method for the estimation of the precision matrix. However, the stepwise procedure requires the known graphical symmetry structure and does not simultaneously perform parameter estimation and model selection.

In the approximate composite likelihood approach, the estimation function is defined from low-dimensional conditional or marginal distributions.
It is typically employed
when full likelihood is computationally infeasible or computationally expensive. Attractive applications of composite likelihood have emerged rapidly to deal with problems with longitudinal data \citep{fieuws2006pairwise}, survival data \citep{parner2001composite}, missing data \citep{yi2011robust}, statistical genetics \citep{larribe2011composite} and Bayesian inference \citep{ribatet2012bayesian}.

This paper proposes a penalized composite likelihood method that performs model selection and precision matrix estimation simultaneously in colored graphical Gaussian models. We employ $L_{1}$ penalties on the off-diagonal elements and on the difference of pairwise elements of the precision matrix. This idea is motivated by simultaneous grouping and feature selection in linear regression \citep{Shen12}. The $L_{1}$ penalties encourage sparsity and, at the same time, give symmetric estimates of the precision matrix. In addition, we develop a computationally efficient method using the difference of convex functions (DC) algorithm, the augmented Lagrangian approach and the coordinate descent optimization. By combining composite likelihood, we further derive a strategy to convert a matrix optimization problem into several much simpler quadratic problems.

The rest of this paper is organized as follows. Section 2 describes colored graphical models and composite likelihood. Section 3 gives the regularization methodology for estimating the precision matrix in the framework of a colored graphical model. Section 4 presents exhaustive numerical examples to demonstrate the promising performance of the penalized composite likelihood method. Section 5 illustrates an application of the proposed method on a glioblastoma gene expression data set. Section 6 concludes the manuscript with a summary.

\section{Preliminaries}

\subsection{Colored graphical models}

We consider an undirected graph $G = (V,E)$ where $V=\{1,2,\cdots,p\}$ and $E$ are the sets of vertices and undirected edges, respectively.  Let $X = (X_{v}, v \in V)$ be a $p$ dimensional random vector following a multivariate normal distribution $N_{p}(\mu,\Sigma)$. Let $\Theta=(
\theta_{ij})_{p\times p}=\Sigma^{-1}$ be the precision matrix. For an undirected graph $G = (V,E)$, we consider a positive definite cone $P_{G}$ of matrices with the element $\theta_{ij}=0$ whenever the edge $(i,j)\notin E$. A well-known property of graphical Gaussian models $\{N_p(\mu,\Theta^{-1}),
\Theta  \in P_{G}\}$ is $X_i$ is conditionally independent of $X_j$ given all the remaining variables $X_{V\setminus \{i,j \}}$ if and only if $\theta_{ij}=0$. In addition, we can assume $\mu=0$ without any loss of generality. If $\mu \neq 0$, all arguments remain valid by simply centering our data.
%The conditional covariance matrix of $X_{i}$ given all the other variables is $\Sigma_{ii\cdot V\setminus \{i\}} = [(\Sigma^{-1})_{ii}]^{-1}=1/K_{ii}$.

Now, let $\mathcal{V}=\{V_1,\cdots,V_{R}\}$ be a partition of $V$ where all vertices in $V_k, k=1,\cdots,R,$ have the same color. Similarly, let $\mathcal{E} =\{E_1,\cdots,E_S\}$ be a partition of $E$ where all edges in $E_r, r=1,\cdots,S,$ have the same color. We call that
$\mathcal{V}$ and $\mathcal{E}$ are the coloring of vertices and edges of the graph $G$, respectively, and denote $\mathcal{G} = (\mathcal{V}, \mathcal{E})$ as a colored graph. A colored graphical model is defined as a RCON model in \cite{Hoj08} with colored classes $(\mathcal{V},\mathcal{E})$ by restricting the elements of the precision matrix $\Theta$ as follows:
\begin{enumerate}
\item In the same color class of vertices, the corresponding diagonal entries of $\Theta$ are equal.
\item In the same color class of edges, the corresponding off-diagonal elements of $\Theta$ are equal.
\end{enumerate}
For detailed description of colored graphical models, we refer the readers to \cite{Hoj08}.

To the best of our knowledge, this is the first work to perform model selection for colored graphical Gaussian models in the frequentist framework. The main objective of this study is to develop the procedure of model selection for the colored graphical models based on the penalty function and composite likelihood.
\subsection{Composite likelihood}
Composite likelihood is derived by multiplying a sequence of conditional or marginal densities. Maximum composite likelihood is especially useful when the full likelihood is not applicable to compute or analytically unknown. Let $Y=(Y_1,\cdots,Y_p)^\top$ be a random vector with the probability density function $f(y|\theta)$ for a $d$-dimensional unknown parameter $\theta\in R^d, d\geq 1$. Composite likelihood is defined on events $\{A_i; i=1,\cdots,k\}$ in the sample space. As in \cite{lindsay1988composite}, a composite likelihood is defined as
$$L_{c}(\theta) = L_{c}(\theta|y) = \prod^k_{i=1}f(y\in A_i|\theta)^{w_i},$$
where $w_i, i=1,\cdots,k$, are positive weights.

Even though the composite conditional likelihood is a pseudo likelihood, the maximum log composite likelihood procedure can still provide consistent estimation. The reader is referred to \cite{varin2011overview} for good asymptotic properties of composite likelihood. For the colored graphical models, the composite likelihood procedure does not rely on the large matrix inversion and is more flexible for the computations.
\section{Proposed method}
\subsection{Composite likelihood estimation}
Let $X_{-j}$ be all components of $X$ except $X_{j}$.
The conditional distribution of $X_{j}$ given $X_{-j}$ is a univariate normal \citep{Mardia79}
\begin{eqnarray*}
\label{condision}
X_{j}|X_{-j}\sim N\Big(-\sum\limits_{i\neq j}X_{i}\frac{\theta_{ij}}{\theta_{jj}}, \frac{1}{\theta_{jj}}\Big)
\end{eqnarray*}
with the density function
$$f(x_j|\Theta,x_{V\setminus \{j\}}) = \frac{\theta_{jj}^{1/2}}{\sqrt{2\pi}}\exp \big\{-\frac{1}{2}\theta_{jj}[x_j+\theta_{jj}^{-1}(\sum\limits^p_{i=1,i\neq j}\theta_{ji}x_i)]^2\big\}.$$
Let $x_{ki}$ denote the element of the $n\times p$ data matrix $X$ for the $k$th individual. The conditional composite likelihood function is
\begin{eqnarray*}
L_{c}(\Theta)=\prod^{n}_{k=1}\prod_{j=1}^{p}\frac{\theta^{\frac{1}{2}}_{jj}}{\sqrt{2\pi}}\exp\{-\frac{1}{2}\theta_{jj}\big[x_{kj}+\theta^{-1}_{jj}(\sum_{i\neq j}\theta_{ij}x_{ki})\big]^2\}.
\end{eqnarray*}
Therefore, the composite log-likelihood can be written as
\begin{eqnarray*}
l_{c}(\Theta)&=&\sum_{j=1}^{p}\sum^{n}_{k=1}\{\frac{1}{2}\log \theta_{jj}-\frac{1}{2}\theta_{jj}\big[x_{kj}+\theta^{-1}_{jj}(\sum_{i\neq j}\theta_{ij}x_{ki})\big]^2\}\\
&=&\frac{1}{2}\sum_{j=1}^{p}\{n\log \theta_{jj}-\theta_{jj}\sum^{n}_{k=1}\big[x_{kj}+\theta^{-1}_{jj}(\sum_{i\neq j}\theta_{ij}x_{ki})\big]^2\}
\end{eqnarray*}
up to a constant. Let $\alpha_{ij}=-\theta_{ij}/\theta_{jj}$, we rewrite the composite log-likelihood in a matrix format as
\begin{eqnarray*}
l_{c}(\Theta)
&=&\frac{1}{2}\sum_{j=1}^{p}\{n\log \theta_{jj}-\theta_{jj}||X_{(j)}-XB_{j}||^2\},
\end{eqnarray*}
where $B_{j}$ and $X_{(j)}=(x_{1j}, x_{2j}, \cdots, x_{nj})^\top$ are the $j$th columns of the matrix $(\alpha_{ij})_{p\times p}$ except for a zero at the $j$th row, and the matrix $X$, respectively.

In linear regression, \cite{Shen12} proposed a method for simultaneous supervised clustering and feature selection among predictors. They also presented an efficient algorithm to seek a parsimonious model by identifying homogeneous groups of regression coefficients, including the zero group and the similar group of coefficients. The method was achieved by performing the regression subject to penalties which encourage sparsity and similarity in estimated coefficients. We will adopt this idea to model selection of colored graphical models in this manuscript.

We consider model selection of colored graphical models by adding penalty functions on elements of the precision matrix $\Theta$.
%Let $G_{d}=(V_{d},E_{d})$ be a complete graph corresponding to the diagonal entries of $\theta_{ii}$ with $V_{d}=\{1,2, \ldots, p\}$ and $G_{o}=(V_{o},E_{o})$ be a complete graph corresponding to the off diagonal entries of $\theta_{jk}$, $k>j$ with $V_{o}=\{1,2, \ldots, \frac{p(p-1)}{2}\}$. We also denote $A:=\{(j,k); k>j, j=1,2, \ldots, p, k=1,2, \ldots, p\}$ and $B:=\{m; m=1,2,\ldots, \frac{p(p-1)}{2}\}$.
%Let us define a map $\gamma$
%$$(j,k) \mapsto (j-1)p-\sum\limits^{j-1}_{i=0}i+k-j, \hspace{10mm} A \rightarrow B.$$
%Let $\beta_{m}=\theta_{\gamma^{-1}(m)}$, $m=1,2,\ldots,\frac{p(p-1)}{2}$.
%
%Actually, we order $\theta_{ij}, i\neq j$, in lexicographical order.
Let us rewrite the off-diagonal elements $\theta_{ij}$, $i\neq j$, according to the lexicographical order as a vector $\beta = \{\beta_1, \cdots, \beta_{\frac{p(p-1)}{2}}\}^\top$. Rewrite the matrix $\Theta$ as a parameter vector
$\theta=(\theta_{11}, \theta_{22},\cdots,\theta_{pp}, \beta^\top)^\top$.
%We denote $\theta=(\theta_{11}, \theta_{22},\cdots,\theta_{pp}, \beta_{1}, \beta_{2},\cdots, \beta_{\frac{p(p-1)}{2}})$.
We will simply write $l_{c}(\Theta)=l_{c}(\theta)$ and $-l_{c}(\theta)$ is asymptotically convex within the neighborhood of the true parameter value $\theta_0$ \citep{gao2015estimation}.

For model selection of colored graphical models, we propose a regularized minimum approach for the negative log composite likelihood
\begin{eqnarray}
\label{min}
\min_{\theta}f(\theta)&=&\min_{\theta}\big\{-\frac{1}{n}l_{c}(\theta)+\lambda_{1} \sum\limits_{j < j'}J_{\tau}(|\theta_{jj}-\theta_{j'j'}|)+\lambda_{2}\sum\limits_{j=1}^{\frac{p(p-1)}{2}}J_{\tau}(|\beta_{j}|)\nonumber\\
&&+\lambda_{3}\sum\limits_{j < j'}J_{\tau}(|\beta_{j}-\beta_{j'}|)\big\}
\end{eqnarray}
where $J_{\tau}(x)= \min \Big(\frac{x}{\tau},1\Big)$, $\lambda_{1}, \lambda_{2}$ and $\lambda_{3}$ are nonnegative tuning parameters controlling the trade-off between the model fit and symmetry structures, $\tau$ is a threshold parameter determining the strength of penalization on off-diagonal elements and differences between element pairs of the precision matrix. The above four parameters can be tuned efficiently by Bayesian information criterion (BIC) for composite likelihood \citep{gao2010composite}. More details can be found in Sections 4 and 5.

\subsection{Computation}

This section develops a relaxation method to address the problem of minimizing the nonconvex function in the expression \eqref{min} using DC programming \citep{le1997solving}. At each iteration, we solve the convex problem by integrating the augmented Lagrange approach \citep{fortin2000augmented} and the coordinate descent optimization. Let us first decompose $f(\theta)$ into a difference $f_{1}(\theta)-f_{2}(\theta)$ of two convex functions with
\begin{eqnarray*}
f_{1}(\theta)&=&-\frac{1}{n}l_{c}(\theta)+\frac{\lambda_{1}}{\tau}\sum\limits_{j < j'}|\theta_{jj}-\theta_{j'j'}|+\frac{\lambda_{2}}{\tau}\sum\limits_{j=1}^{\frac{p(p-1)}{2}}|\beta_{j}|+\frac{\lambda_{3}}{\tau}\sum\limits_{j < j'}|\beta_{j}-\beta_{j'}|\\
\end{eqnarray*}
and
\begin{eqnarray*}
f_{2}(\theta)&=&\frac{\lambda_{1}}{\tau}\sum\limits_{j < j'}(|\theta_{jj}-\theta_{j'j'}|-\tau)_{+}+\frac{\lambda_{2}}{\tau}\sum\limits_{j=1}^{\frac{p(p-1)}{2}}(|\beta_{j}|-\tau)_{+}+\frac{\lambda_{3}}{\tau}\sum\limits_{j < j'}(|\beta_{j}-\beta_{j'}|-\tau)_{+}.\\
\end{eqnarray*}
Here $y_{+}$ is the positive part of $y$. Next, we approximate the convex function $f_{2}(\theta)$ iteratively by its piecewise affine minorization, that is, at iteration $m$,
\begin{eqnarray*}
&&f_{2}(\hat{\theta}^{(m-1)})+\frac{\lambda_{1}}{\tau}\sum\limits_{j < j'}I(|\hat{\theta}^{(m-1)}_{jj}-\hat{\theta}^{(m-1)}_{j'j'}|\geq \tau)(|\theta_{jj}-\theta_{j'j'}|-|\hat{\theta}^{(m-1)}_{jj}-\hat{\theta}^{(m-1)}_{j'j'}|)\\
&&\hspace{4mm}+\frac{\lambda_{2}}{\tau}\sum\limits_{j=1}^{\frac{p(p-1)}{2}}I(|\hat{\beta}^{(m-1)}_{j}|\geq \tau)(|\beta_{j}|-|\hat{\beta}^{(m-1)}_{j}|)\\
&& \hspace{4mm}+\frac{\lambda_{3}}{\tau}\sum\limits_{j < j'}I(|\hat{\beta}^{(m-1)}_{j}-\hat{\beta}^{(m-1)}_{j'}|\geq \tau)(|\beta_{j}-\beta_{j'}|-|\hat{\beta}^{(m-1)}_{j}-\hat{\beta}^{(m-1)}_{j'}|).
\end{eqnarray*}
This approach leads to an upper convex approximation function
\begin{eqnarray}\label{mincov}
f^{(m)}(\theta)&=&-\frac{1}{n}l_{c}(\theta)+\frac{\lambda_{1}}{\tau}\sum_{j<j':(j,j')\in E_{d}^{(m-1)}} |\theta_{jj}-\theta_{j'j'}|+\frac{\lambda_{2}}{\tau}\sum_{j:j\in V_{o}^{(m-1)}} |\beta_{j}|\nonumber\\
&&
+\frac{\lambda_{3}}{\tau}\sum_{j<j':(j,j')\in E_{o}^{(m-1)}} |\beta_{j}-\beta_{j'}|
\end{eqnarray}
where
$$E_{d}^{(m-1)}=\{j<j'; |\hat{\theta}^{(m-1)}_{jj}-\hat{\theta}^{(m-1)}_{j'j'}|<\tau\},$$
$$V_{o}^{(m-1)}=\{j\in (1,2,\ldots, \frac{p(p-1)}{2}); |\hat{\beta}^{(m-1)}_{j}|< \tau\}$$
and
$$E_{o}^{(m-1)}=\{j<j'; |\hat{\beta}^{(m-1)}_{j}-\hat{\beta}^{(m-1)}_{j'}|<\tau\}.$$

To optimize the $m$th iteration \eqref{mincov}, we finally use an iterative approach based on the augmented Lagrange method and the coordinate descent optimization. We first define new variables $k_{jj'}=\theta_{jj}-\theta_{j'j'}$ and $\beta_{jj'}=\beta_{j}-\beta_{j'}$ for $j \neq j'$. Let
\begin{eqnarray*}
\xi=\Big(\theta_{11}, \theta_{22}, \ldots, \theta_{pp}, k_{12}, k_{13}, \ldots, k_{1p}, k_{23}, k_{24}, \ldots, k_{2p}, \ldots, k_{(p-1)p},\beta_{1}, \beta_{2}, \ldots, \beta_{\frac{p(p-1)}{2}},\\
 \beta_{12}, \beta_{13}, \ldots, \beta_{1\frac{p(p-1)}{2}}, \beta_{23}, \beta_{24}, \ldots, \beta_{2\frac{p(p-1)}{2}}, \ldots, \beta_{\frac{p(p-1)-2}{2}\frac{p(p-1)}{2}}\Big).
\end{eqnarray*}
Minimizing the expression \eqref{mincov} is then equivalent to minimize the following function
\begin{eqnarray}\label{addvariables}
\tilde{f}^{(m)}(\xi)&=&-\frac{1}{n}l_{c}(\theta)+\frac{\lambda_{1}}{\tau}\sum_{j<j':(j,j')\in E_{d}^{(m-1)}} |k_{jj'}|+\frac{\lambda_{2}}{\tau}\sum_{j:j\in V_{o}^{(m-1)}} |\beta_{j}|\nonumber\\
&&
+\frac{\lambda_{3}}{\tau}\sum_{j<j':(j,j')\in E_{o}^{(m-1)}} |\beta_{jj'}|.
\end{eqnarray}
For the $m$th iteration, we use the augmented Lagrange algorithm to minimize the function \eqref{addvariables} iteratively with respect to $t$. At the $t$th iteration, we minimize
\begin{eqnarray}\label{augmented}
\bar{f}^{(m)}(\xi)&=&\tilde{f}^{(m)}(\xi)+\sum_{j<j':(j,j')\in E_{d}^{(m-1)}}a^{(t)}_{jj'} (\theta_{jj}-\theta_{j'j'}-k_{jj'})\nonumber\\
&&\hspace{4mm}+\frac{1}{2}\sum_{j<j':(j,j')\in E_{d}^{(m-1)}}b^{(t)}_{jj'} (\theta_{jj}-\theta_{j'j'}-k_{jj'})^2\nonumber\\
&&\hspace{4mm}+\sum_{j<j':(j,j')\in E_{o}^{(m-1)}}c^{(t)}_{jj'} (\beta_{j}-\beta_{j'}-\beta_{jj'})\nonumber\\
&&\hspace{4mm}+\frac{1}{2}\sum_{j<j':(j,j')\in E_{o}^{(m-1)}}d^{(t)}_{jj'} (\beta_{j}-\beta_{j'}-\beta_{jj'})^2
\end{eqnarray}
where $a^{(t)}_{jj'}$, $b^{(t)}_{jj'}$, $c^{(t)}_{jj'}$ and $d^{(t)}_{jj'}$ are Lagrange multipliers. Let $$a^{(t+1)}_{jj'}=a^{(t)}_{jj'}+b^{(t)}_{jj'}(\hat{\theta}^{(m,t)}_{jj}-\hat{\theta}^{(m,t)}_{j'j'}-\hat{k}^{(m,t)}_{jj'}),\hspace{10mm} b^{(t+1)}_{jj'}=\rho b^{(t)}_{jj'},$$
$$c^{(t+1)}_{jj'}=c^{(t)}_{jj'}+d^{(t)}_{jj'}(\hat{\beta}^{(m,t)}_{j}-\hat{\beta}^{(m,t)}_{j'}-\hat{\beta}^{(m,t)}_{jj'}) \hspace{5mm} \text{and} \hspace{5mm} d^{(t+1)}_{jj'}=\rho d^{(t)}_{jj'}$$ where $\rho >1$.
We next use the coordinate descent method to compute $\hat{\xi}^{(m,t)}$ from \eqref{augmented}. For each component of $\xi$, we fix the other components at their current values. The first order derivatives of $\bar{f}^{(m)}(\xi)$ for different components are derived:\\
For $j=1,2,\ldots,p$, we have that
\begin{eqnarray}\label{non1}
\frac{\partial \bar{f}^{(m)}(\xi)}{\partial \theta_{jj}}
%&=&\frac{1}{2n}(-n\theta^{-1}_{jj}+X^T_{(j)}X_{(j)}-\theta^{-2}_{jj}\sum\limits_{k\neq j}\sum\limits_{l\neq j}\theta_{kj}\theta_{lj}X^T_{(k)}X_{(l)})+\sum_{j<j':(j,j')\in E_{d}^{(m-1)}}a^{(t)}_{jj'}\nonumber\\
%&&-\sum_{j'<j:(j,j')\in E_{d}^{(m-1)}}a^{(t)}_{j'j}+\sum_{j<j':(j,j')\in E_{d}^{(m-1)}}b^{(t)}_{jj'} (\theta_{jj}-\theta_{j'j'}-k_{jj'})\nonumber\\
%&&-\sum_{j'<j:(j,j')\in E_{d}^{(m-1)}}b^{(t)}_{j'j} (\theta_{j'j'}-\theta_{jj}-k_{j'j})\nonumber\\
&=&-\frac{1}{2n}\theta^{-2}_{jj}\sum\limits_{k\neq j}\sum\limits_{l\neq j}\theta_{kj}\theta_{lj}X^\top_{(k)}X_{(l)}-\frac{1}{2}\theta^{-1}_{jj}+\sum_{j<j':(j,j')\in E_{d}^{(m-1)}}b^{(t)}_{jj'}\theta_{jj} \nonumber\\
&&+\sum_{j'<j:(j,j')\in E_{d}^{(m-1)}}b^{(t)}_{j'j}\theta_{jj}+\frac{1}{2n}X^\top_{(j)}X_{(j)}+\sum_{j<j':(j,j')\in E_{d}^{(m-1)}}a^{(t)}_{jj'}\nonumber\\
&&-\sum_{j'<j:(j,j')\in E_{d}^{(m-1)}}a^{(t)}_{j'j}-\sum_{j<j':(j,j')\in E_{d}^{(m-1)}}b^{(t)}_{jj'} (\theta_{j'j'}+k_{jj'})\nonumber\\
&&-\sum_{j'<j:(j,j')\in E_{d}^{(m-1)}}b^{(t)}_{j'j} (\theta_{j'j'}-k_{j'j}).
\end{eqnarray}
Let $(q,l)$ be the index in the matrix $\Theta$ corresponding to the index $j$ in the vector $\beta$ according to the lexicographical order. For $j\in V^{(m-1)}_{o}$, we have that
%$(q,l)=\gamma^{-1}(j)$
\begin{eqnarray}\label{non2}
\frac{\partial \bar{f}^{(m)}(\xi)}{\partial \beta_{j}}&=&\frac{1}{n}\Big(\theta^{-1}_{qq}X^\top_{(l)}X_{(l)}\theta_{ql}+X^\top_{(q)}X_{(l)}+\theta^{-1}_{qq}\sum\limits_{i\notin \{q,l\}}X^\top_{(l)}X_{(i)}\theta_{iq}+\theta^{-1}_{ll}X^\top_{(q)}X_{(q)}\theta_{ql}\nonumber\\
&&+X^\top_{(l)}X_{(q)}+\theta^{-1}_{ll}\sum\limits_{i\notin \{q,l\}}X^\top_{(q)}X_{(i)}\theta_{il}\Big)+\frac{\lambda_{2}}{\tau}\frac{d|\beta_{j}|}{d\beta_{j}}+\sum_{j<j':(j,j')\in E_{o}^{(m-1)}}c^{(t)}_{jj'}\nonumber\\
&&-\sum_{j'<j:(j,j')\in E_{o}^{(m-1)}}c^{(t)}_{j'j}+\sum_{j<j':(j,j')\in E_{o}^{(m-1)}}d^{(t)}_{jj'} (\beta_{j}-\beta_{j'}-\beta_{jj'})\nonumber\\
&&-\sum_{j'<j:(j,j')\in E_{o}^{(m-1)}}d^{(t)}_{j'j} (\beta_{j'}-\beta_{j}-\beta_{j'j}).
\end{eqnarray}
For $j\notin V^{(m-1)}_{o}$, the partial derivative with respect to $\beta_{j}$ is
\begin{eqnarray}\label{non3}
\frac{\partial \bar{f}^{(m)}(\xi)}{\partial \beta_{j}}&=&\frac{1}{n}\Big(\theta^{-1}_{qq}X^\top_{(l)}X_{(l)}\theta_{ql}+X^\top_{(q)}X_{(l)}+\theta^{-1}_{qq}\sum\limits_{i\notin \{q,l\}}X^\top_{(l)}X_{(i)}\theta_{iq}+\theta^{-1}_{ll}X^\top_{(q)}X_{(q)}\theta_{ql}\nonumber\\
&&+X^\top_{(l)}X_{(q)}+\theta^{-1}_{ll}\sum\limits_{i\notin \{q,l\}}X^\top_{(q)}X_{(i)}\theta_{il}\Big)+\sum_{j<j':(j,j')\in E_{o}^{(m-1)}}c^{(t)}_{jj'}\nonumber\\
&&-\sum_{j'<j:(j,j')\in E_{o}^{(m-1)}}c^{(t)}_{j'j}+\sum_{j<j':(j,j')\in E_{o}^{(m-1)}}d^{(t)}_{jj'} (\beta_{j}-\beta_{j'}-\beta_{jj'})\nonumber\\
&&-\sum_{j'<j:(j,j')\in E_{o}^{(m-1)}}d^{(t)}_{j'j} (\beta_{j'}-\beta_{j}-\beta_{j'j}).
\end{eqnarray}
For $(j,j')\in E_{d}^{(m-1)}$, the following equality holds
\begin{eqnarray}\label{lin1}
\frac{\partial \bar{f}^{(m)}(\xi)}{\partial k_{jj'}}&=&\frac{\lambda_{1}}{\tau}\frac{d|k_{jj'}|}{dk_{jj'}}-a^{(t)}_{jj'}
-b^{(t)}_{jj'} (\theta_{jj}-\theta_{j'j'}-k_{jj'}).
\end{eqnarray}
For $(j,j')\in E_{o}^{(m-1)}$, we can compute that
\begin{eqnarray}\label{lin2}
\frac{\partial \bar{f}^{(m)}(\xi)}{\partial \beta_{jj'}}&=&\frac{\lambda_{3}}{\tau}\frac{d|\beta_{jj'}|}{d\beta_{jj'}}-c^{(t)}_{jj'}
-d^{(t)}_{jj'} (\beta_{jj}-\beta_{j'j'}-\beta_{jj'}).
\end{eqnarray}
Let us continue to solve this problem by setting \eqref{non1}, \eqref{non2}, \eqref{non3}, \eqref{lin1} and \eqref{lin2} equal to 0 which results in five equations. The first equation is a one variable cubic equation. We can find its positive real root through Van Wijngaarden-Dekker-Brent method \citep{brent2013algorithms}. In implementation, we use the function uniroot.all() in R package rootSolve. The roots for the last four equations are given as follows:\\
For $j\in V^{(m-1)}_{o}$, $\frac{\partial \bar{f}^{(m)}(\xi)}{\partial \beta_{j}}=0$ implies
\begin{eqnarray*}
\hat{\beta}^{(m,t)}_{j}&=&ST(\frac{\hat{z}^{(m,t)}_{j}}{\hat{r}^{(m,t)}_{j}},\frac{\lambda_{2}}{\tau \hat{r}^{(m,t)}_{j}}).
\end{eqnarray*}
For $j\notin V^{(m-1)}_{o}$, if $\frac{\partial \bar{f}^{(m)}(\xi)}{\partial \beta_{j}}=0$, then
\begin{eqnarray*}
\hat{\beta}^{(m,t)}_{j}&=&\frac{\hat{z}^{(m,t)}_{j}}{\hat{r}^{(m,t)}_{j}}.
\end{eqnarray*}
Here
\begin{eqnarray*}
\hat{r}^{(m,t)}_{j} &=& \frac{1}{n}(\hat{\theta}_{qq}^{(m,t)})^{-1}X^\top_{(l)}X_{(l)}+\frac{1}{n}(\hat{\theta}_{ll}^{(m,t)})^{-1}X^\top_{(q)}X_{(q)}+\sum_{j<j':(j,j')\in E_{o}^{(m-1)}}d^{(t)}_{jj'}\\
&&+\sum_{j'<j:(j,j')\in E_{o}^{(m-1)}}d^{(t)}_{j'j}
\end{eqnarray*}
and
%$a_{j} = a_{ql}$ b_{j} = b_{ql}
\begin{eqnarray*}
\hat{z}^{(m,t)}_{j} &=& -\frac{1}{n}X^\top_{(q)}X_{(l)}-\frac{1}{n}(\hat{\theta}^{(m,t)}_{qq})^{-1}\sum\limits_{i\notin \{q,l\}}X^\top_{(l)}X_{(i)}\hat{\theta}^{(m,t)}_{iq}-\frac{1}{n}X^\top_{(l)}X_{(q)}\\
&& -\frac{1}{n}(\hat{\theta}_{ll}^{(m,t)})^{-1}\sum\limits_{i\notin \{q,l\}}X^\top_{(q)}X_{(i)}\hat{\theta}^{(m,t)}_{il}-\sum_{j<j':(j,j')\in E_{o}^{(m-1)}}c^{(t)}_{jj'}+\sum_{j'<j:(j,j')\in E_{o}^{(m-1)}}c^{(t)}_{j'j}\\
&&+\sum_{j<j':(j,j')\in E_{o}^{(m-1)}}d^{(t)}_{jj'} (\hat{\beta}^{(m,t)}_{j'}+\hat{\beta}^{(m,t)}_{jj'})+\sum_{j'<j:(j,j')\in E_{o}^{(m-1)}}d^{(t)}_{j'j} (\hat{\beta}^{(m,t)}_{j'}-\hat{\beta}^{(m,t)}_{j'j}).
\end{eqnarray*}
For $(j,j')\in E_{d}^{(m-1)}$, solving $\frac{\partial \bar{f}^{(m)}(\xi)}{\partial k_{jj'}}=0$ gives
\begin{eqnarray*}
\hat{k}^{(m,t)}_{jj'}&=&ST\Big(\frac{a^{(t)}_{jj'}+b^{(t)}_{jj'}(\hat{\theta}^{(m,t)}_{jj}-\hat{\theta}^{(m,t)}_{j'j'})}{b^{(t)}_{jj'}},\frac{\lambda_{1}}{\tau b^{(t)}_{jj'}}\Big).
\end{eqnarray*}
For $(j,j')\in E_{o}^{(m-1)}$, $\frac{\partial \bar{f}^{(m)}(\xi)}{\partial \beta_{jj'}}=0$ implies
\begin{eqnarray*}
\hat{\beta}^{(m,t)}_{jj'}&=&ST\Big(\frac{c^{(t)}_{jj'}+d^{(t)}_{jj'}(\hat{\beta}^{(m,t)}_{j}-\hat{\beta}^{(m,t)}_{j'})}{d^{(t)}_{jj'}},\frac{\lambda_{3}}{\tau d^{(t)}_{jj'}}\Big).
\end{eqnarray*}
Here $ST(z,\gamma)=sign(z)(|z|-\gamma)_{+}$ is the soft thresholding operator. This whole process of coordinate descent is repeated iteratively until it converges.

\section{Statistical performance}

\begin{figure}[htbp]
    \hspace{-4mm}
    \begin{minipage}{0.3\linewidth}
        \centering
        \includegraphics[width=1.55in]{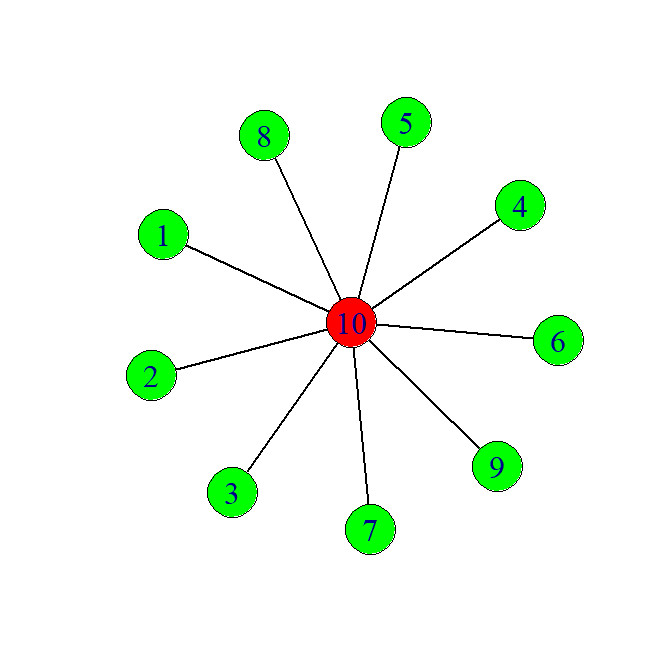}\\
        (a)
    \end{minipage}
    \begin{minipage}{0.3\linewidth}
        \centering
        \includegraphics[width=2in]{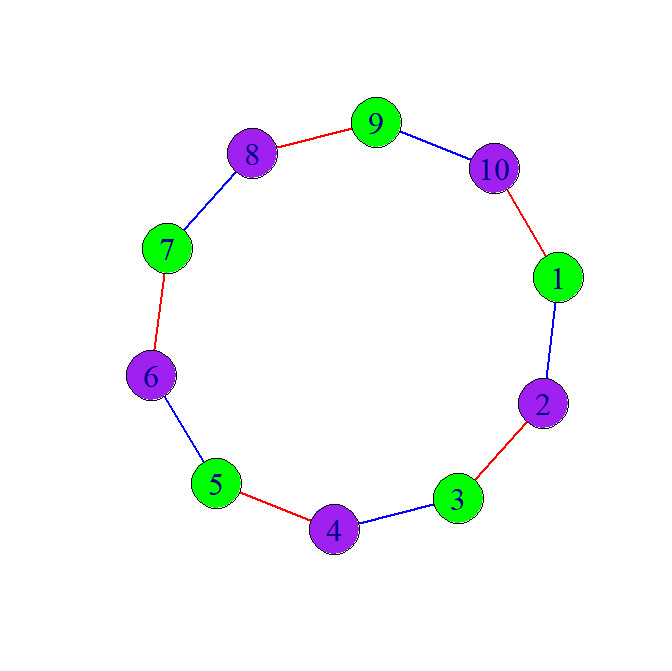}\\
        (b)
    \end{minipage}
    \begin{minipage}{0.3\linewidth}
        \centering
        \includegraphics[width=2in]{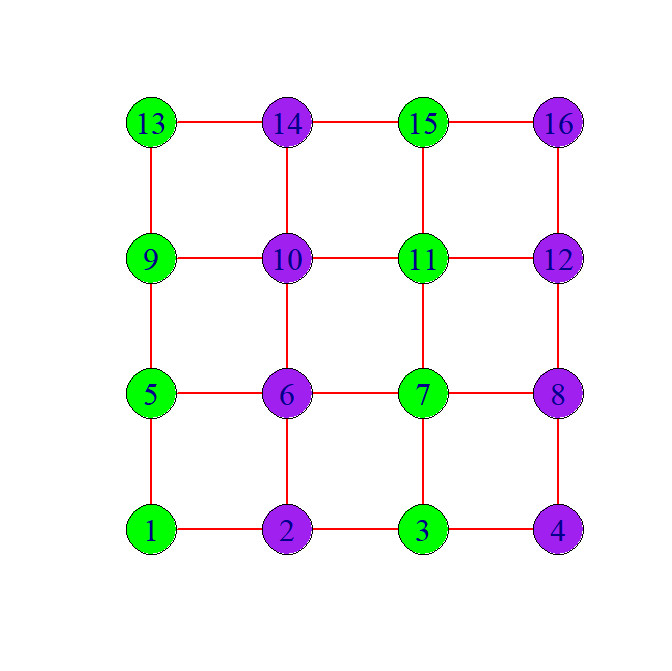}\\
        (c)
    \end{minipage}
    \caption{(a) Star graph with $p=10$; (b) Cycle graph with $p=10$; (c) Grid graph with $4\times 4$ vertices.}
    \label{fig:1}
\end{figure}

We begin by illustrating the performance of the proposed method on three types of colored graphical models underlying the colored graphs: stars, cycles and grid graphs as displayed in Figure \ref{fig:1}. We simulate data sets with the number of parameters $p$ ranging from 10 to 30, each with a sample size $n$ ranging from 250 to 1000.  These values are used by all the 100 simulated datasets. The tuning parameters $\lambda_i, i=1,2,3,$ and the threshold parameter $\tau$ are selected based on composite likelihood BIC with $BIC_{c} = -2 l_c(\hat{\theta})+ df \log n$. Here $df$ denotes the total number of parameters in the precision matrix \citep{gao2010composite}. The parameters $\lambda_i, i=1,2,3,$ and $\tau$ are obtained by minimizing $BIC_{c}$ in a four-dimensional parameter space using a grid search procedure.
Performance metrics are provided to measure the accuracy of the precision matrix estimation as well as that of the identification of the zero and symmetry structures. Let $\Theta^{T}$ and $\hat{\Theta}$ be the true precision matrix and the selected precision matrix through the model selection approach, respectively. To evaluate the performance of an estimator $\hat{\Theta}$, we use the empirical normalized mean squared error \citep{meng2014marginal} defined as
%entropy loss function, also known as the Kullback-Leibler loss (KLL) function, defined as follows:
%$$KLL(\hat{\Theta},\Theta) = trace(\Theta^{-1}\hat{\Theta})-\log det(\Theta^{-1}\hat{\Theta})-p.$$
%The KLL function has been used widely in the prior research on estimation of covariance and concentration matrices \citep[see, e.g.][]{yuan2007model,rothman2008sparse,yin2013adjusting}. Moreover, we also use the mean squared error defined as
$$MSE(\hat{\Theta},\Theta)=\frac{||\hat{\Theta}-\Theta||^2_2}{||\Theta^2_2||}.$$

Regarding the sparsity pattern, we use the $F_1$-score \citep{mohammadi2015bayesian} to evaluate the performance defined as
$$F_1-\text{score}= \frac{2TP}{2TP+FP+FN}$$
where $TP$, $FP$, and $FN$ are the number of correctly estimated nonzero entries, the number of incorrectly estimated nonzero entries and the number of incorrectly estimated zero entries, respectively. The $F_1$-score is ranging from 0 to 1. The value 1 is attributed to perfect performance.

%we compute specificity, sensitivity, and Matthews correlation coefficient (MCC), defined as
%$$Specificity = TN/(TN+FP), \ \ \ \ \ \ \ \  Sensitivity = TP/(TP+FN),$$
%$$MCC=\frac{TP\times TN-FP\times FN}{\sqrt{(TP+FP)(TP+FN)(TN+FP)(TN+FN)}}$$
%where TP, TN, FP, and FN are the number of correctly estimated nonzero entries, the number of correctly estimated zero entries, the number of incorrectly estimated nonzero entries and the number of incorrectly estimated zero entries, respectively.

To assess the performance of the symmetry structure, we define $B$ as a set of edges in which the corresponding $\Theta^{T}_{ij}=0$ if $(i,j) \in B$. Let $V^{T}_{i}$, $i=1,\cdots,t$, be the vertex color class and $E^{T}_{j}$, $j=1,\cdots,k$, be the edge color class of the true graph. We use the measures $d_{0}$, $d_{V^{T}_{i}}$, $i=1,\cdots,t$, $d_{E^{T}_{j}}$, $j=1,\cdots,k$, and $Acc_{all}$ for measuring supervised clustering and feature selection in \cite{Shen12}. We define
$$d_{0} = \{\sum\limits_{(i,j)\in B}1_{\hat{\Theta}_{ij}=0}+\sum\limits_{(i,j)\notin B}1_{\hat{\Theta}_{ij}\neq 0}\}/\frac{p(p-1)}{2}$$
which measures the performance in identifying zero constraints. For $i=1,\cdots,t$, let
$$d_{V^{T}_{i}} = \frac{\sum\limits_{j\neq j':(j,j)\in V^{T}_{i},(j',j')\in V^{T}_{i}}1_{\hat{\Theta}_{jj}=\hat{\Theta}_{j'j'}}+\sum\limits_{j\neq j':(j,j)\in V^{T}_{i},(j',j')\notin V^{T}_{i}}1_{\hat{\Theta}_{jj}\neq \hat{\Theta}_{j'j'}}}{|V^{T}_{i}|(p-1)}$$
which measures the performance in identifying the true vertex color classes. For $j=1,\cdots,k$, let
$$d_{E^{T}_{j}} = \frac{\sum\limits_{\substack{(k,l)\neq (k',l'):\\(k,l)\in E^{T}_{j},\\(k',l')\in E^{T}_{j}}}1_{\hat{\Theta}_{kl}=\hat{\Theta}_{k'l'}}+\sum\limits_{\substack{(k,l)\neq (k',l'):\\(k,l)\in E^{T}_{j},\\(k',l')\notin E^{T}_{j}}}1_{\hat{\Theta}_{kl}\neq \hat{\Theta}_{k'l'}}}{|E^{T}_{j}|[\frac{p(p-1)}{2}-1]}$$
which measures the performance in identifying the true edge color classes. We further define
$$Acc_{all} = \frac{d_{0}+\sum\limits^{s}_{i=1}d_{V^{T}_{i}}+\sum\limits^{t}_{j=1}d_{E^{T}_{j}}}{1+s+t}.$$ Note that $Acc_{all}$ lies between 0 and 1. A better identification of the true colored model is associated with a larger value of $Acc_{all}$.

We simulate samples from the multivariate normal $N(0, (\Theta^T)^{-1})$. Corresponding to different sparsities and symmetry patterns, we consider the following three different kinds of colored graphs:
%If p=10, the number of estimated parameter is p(p+1)/2= 55. If p=20, the number of estimated parameters is 210. If p=30, the number of estimated parameters is 465. If we have p=31, then the number of parameters is 496.
\begin{enumerate}
\item
Star graphs: For the star graph with $p$ vertices, the edge set is $E=\{(i,p):1\leq i\leq p-1\}$. Let $\Theta^T_{ii} = 1$($1\leq i\leq p-1$), $\Theta^T_{pp} = 2$, $\Theta^T_{ip} = \Theta^T_{pi} = 0.25$($1\leq i\leq p-1$), and $\Theta^T_{ij}=0$ if $(i,j) \notin E$. The colored star graph with 10 vertices is shown in Figure \ref{fig:1} (a).

\item
Cycle graphs: For the cycle graph with $p$ vertices, the edge set is $E=\{(i,i+1):1\leq i\leq p-1\}\cup (1,p)$. Let $\Theta^T_{ii} = 1$ if $i$ is odd, $\Theta^T_{ii} = 1.5$ if $i$ is even, $\Theta^T_{ij} = \Theta^T_{ji} = 0.5$ if $i-j=1$ and $i$ is odd, $\Theta^T_{ij} = \Theta^T_{ji} = 0.3$ if $i-j=1$ and $i$ is even, and $\Theta^T_{ij}=0$ if $(i,j) \notin E$. The colored cycle graph with 10 vertices is shown in Figure \ref{fig:1} (b).
\item
Grid graphs: For the grid graph with $p=q*q$ vertices, let $\Theta^T_{ii} = 3$ if $i$ is odd, $\Theta^T_{ii} = 5$ if $i$ is even, $\Theta^T_{ij} = \Theta^T_{ji} = 0.8$ if $(i,j) \in E$, and $\Theta^T_{ij}=0$ if $(i,j) \notin E$. The colored grid graph with $4\times 4$ vertices is shown in Figure \ref{fig:1} (c).
\end{enumerate}

%\begin{table}[h!]
%\centering
%\caption{Summary of performance measures for star graphs.}
%\begin{tabular}{cccccc}
%     \hline
%     \hline
%     $p$ & $ n$ & MSE &$F_1$-score & $d_0$ &$Acc_{all}$  \\ \hline
%          &  250 & 0.050(0.0203) & 0.796(0.2081) & 0.886(0.1794) &0.701(0.0482) \\
%     10   &  500 & 0.030(0.0143) & 0.946(0.1094)& 0.973(0.0896)&0.993(0.0224)\\
%          &  1000 & 0.021(0.0055) & 0.997(0.0129) & 0.999(0.0049)&1.000(0.0012) \\
%     \hline
%          &  250 & 0.142(0.0349) &0.806(0.1128)&0.967(0.0162) & 0.859(0.0091)  \\
%     20   &  500 & 0.103(0.0187) &0.949(0.0392) &0.991(0.0069) &0.969(0.0017)  \\
%          &  1000 & 0.077(0.0074) &0.993(0.0137) &0.999(0.0026) &1.000(0.0007)  \\
%     \hline
%          &  250 & 0.283(0.151) &0.817(0.142)&0.962(0.088) &0.879(0.022)  \\
%     30   &  500 & 0.331(0.178) &0.964(0.087) &0.988(0.083) &0.970(0.021)  \\
%          &  1000 &0.400(0.086) &1.000(0.002) &1.000(0.000) &0.997(0.020)  \\
%     \hline
%   \end{tabular}
%\label{table:1}
%\end{table}

\begin{table}[h!]
\centering
\caption{Summary of performance measures for star graphs.}
\begin{tabular}{cccccc}
     \hline
     \hline
     $p$ & $ n$ & MSE &$F_1$-score & $d_0$ &$Acc_{all}$  \\ \hline
          &  250 & 0.0616(0.0334)& 0.7378(0.2932) & 0.9289(0.0605) &0.9098(0.0227) \\
     10   &  500 & 0.0361(0.0264) & 0.8964(0.2417)& 0.9731(0.0508)&0.9924(0.0150)\\
          &  1000 & 0.0206(0.0051) & 0.9976(0.0116) & 0.9991(0.0044)&0.9998(0.0011) \\
     \hline
          &  250 & 0.2515(0.0726) &0.6154(0.2783)&0.9493(0.0263) & 0.9440(0.0103)  \\
     20   &  500 & 0.1671(0.0600) &0.8924(0.1849) &0.9837(0.0196) &0.9811(0.0065)  \\
          &  1000 & 0.1232(0.0164) &0.9924(0.0152) &0.9985(0.0029) &0.9995(0.0016)  \\
     \hline
          &  250 & 0.9684(0.2135) &0.7934(0.1044)&0.9776(0.0091) &0.9145(0.0025)  \\
     30   &  500 & 0.9032(0.1221) &0.9460(0.0301) &0.9933(0.0036) &0.9633(0.0014)  \\
          &  1000 &0.8582(0.0610) &0.9947(0.0092) &0.9993(0.0012) &0.9997(0.0001)  \\
     \hline
   \end{tabular}
\label{table:1}
\end{table}

%\begin{table}[h!]
%\centering
%\caption{Summary of performance measures for cycle graphs.}
%\begin{tabular}{cccccc}
%     \hline
%     \hline
%     $p$ & $ n$ & MSE &$F_1$-score & $d_0$ &$Acc_{all}$  \\ \hline
%          &  250 & 0.028(0.0141) & 0.884(0.1613) &  0.912(0.1735) &0.920(0.0360) \\
%     10   &  500 & 0.014(0.0062) & 0.977(0.0856)&  0.982(0.0853) &0.959(0.0176)\\
%          &  1000 & 0.010(0.0034) & 0.994(0.0514) & 0.995(0.0467)&0.948(0.0107)\\
%     \hline
%          &  250 & 0.042(0.0163) &0.904(0.1079) &0.971(0.0781)  &0.900(0.0156) \\
%     20   &  500 & 0.030(0.0116) &0.999(0.0035) &1.000(0.0007)  & 0.925(0.0076) \\
%          &  1000 & 0.029(0.0116) &0.999(0.0035) &1.000(0.0007)  &0.981(0.0036) \\
%     \hline
%          &  250 & 0.064(0.0168) & 0.907(0.0467)&0.987(0.0065)  & 0.836(0.0046) \\
%     30   &  500 & 0.056(0.0141) &0.986(0.0169) &0.998(0.0023)  &0.970(0.0030) \\
%          &  1000 & 0.057(0.0130) &1.000(0.0023) &1.000(0.0003) & 0.933(0.0066)  \\
%     \hline
%   \end{tabular}
%\label{table:2}
%\end{table}

\begin{table}[h!]
\centering
\caption{Summary of performance measures for cycle graphs.}
\begin{tabular}{cccccc}
     \hline
     \hline
     $p$ & $ n$ & MSE &$F_1$-score & $d_0$ &$Acc_{all}$  \\ \hline
          &  250 & 0.0237(0.0097) & 0.9123(0.1276) &  0.9436(0.1152) &0.9185(0.0243) \\
     10   &  500 & 0.0123(0.0058) & 0.9770(0.0744)&  0.9838(0.0803) &0.9242(0.0173)\\
          &  1000 & 0.0072(0.0035) & 0.9962(0.0130) & 0.9982(0.0061)&0.9471(0.0068)\\
     \hline
          &  250 & 0.0359(0.0094) &0.9144(0.0541) &0.9821(0.0111)  &0.8643(0.0024) \\
     20   &  500 & 0.0244(0.0063) &0.9898(0.0178) &0.9978(0.0037)  & 0.8877(0.0071) \\
          &  1000 & 0.0223(0.0049) &0.9991(0.0074) &0.9998(0.0017)  &0.9396(0.0009) \\
     \hline
          &  250 & 0.0650(0.0160) & 0.8984(0.0496)&0.9866(0.0060)  & 0.8646(0.0015) \\
     30   &  500 & 0.0575(0.0153) &0.9833(0.0194) &0.9977(0.0026)  &0.9279(0.0067) \\
          &  1000 & 0.0571(0.0130) &0.9997(0.0023) &1.0000(0.0003) & 0.9326(0.0066)  \\
     \hline
   \end{tabular}
\label{table:2}
\end{table}

%\begin{table}[h!]
%\centering
%\caption{Summary of performance measures for grid graphs.}
%\begin{tabular}{cccccc}
%     \hline
%     \hline
%     $p$ & $ n$ & MSE &$F_1$-score & $d_0$ &$Acc_{all}$  \\ \hline
%          &  250   & 0.031(0.0124) & 0.777(0.1351) & 0.785(0.1709)&0.717(0.0434)  \\
%     $3\times 3$   &  500 & 0.011(0.0076) & 0.938(0.0956)& 0.948(0.0970) &0.941(0.0248)\\
%          &  1000   & 0.003(0.0038) & 0.995(0.0148) & 0.997(0.0099) &0.953(0.0054)\\
%     \hline
%          &  250 & 0.045(0.0146) &0.582(0.1771)& 0.658(0.2266)  &0.662(0.0570) \\
%     $4\times 4$   &  500 & 0.017(0.0095) &0.878(0.1790) &0.909(0.1901)  &0.964(0.0490) \\
%          &  1000 & 0.005(0.0040) &0.984(0.0199) &0.994(0.0076) &0.986(0.0024)  \\
%     \hline
%          &  250 & 0.052(0.0204) &0.496(0.2123)&0.629(0.2637)& 0.717(0.0656)   \\
%     $5\times 5$  &  500 & 0.017(0.0060) &0.944(0.0463) &0.985(0.0168)  &0.960(0.0135) \\
%          &  1000 &  0.006(0.003) &0.987(0.015) & 0.997(0.004)   &0.991(0.002)\\
%     \hline
%   \end{tabular}
%\label{table:3}
%\end{table}

\begin{table}[h!]
\centering
\caption{Summary of performance measures for grid graphs.}
\begin{tabular}{cccccc}
     \hline
     \hline
     $p$ & $ n$ & MSE &$F_1$-score & $d_0$ &$Acc_{all}$  \\ \hline
          &  250   & 0.0518(0.0362) & 0.6192(0.3756) & 0.8469(0.1200)&0.8669(0.0307)  \\
     $3\times 3$   &  500 & 0.0228(0.0324) & 0.8243(0.3247)& 0.9294(0.1092) &0.8841(0.0292)\\
          &  1000   & 0.0038(0.0139) & 0.9765(0.1409) & 0.9911(0.0474) &0.9978(0.0118)\\
     \hline
          &  250 & 0.0550(0.0243) &0.6514(0.2293)& 0.9026(0.0459)  &0.9158(0.0160) \\
     $4\times 4$   &  500 & 0.0238(0.0252) &0.8473(0.2313) &0.9563(0.0484)  &0.9746(0.0122) \\
          &  1000 & 0.0055(0.0054) &0.9766(0.0293) &0.9912(0.0107) &0.9978(0.0027)  \\
     \hline
          &  250 & 0.0630(0.0176) &0.6471(0.1513)&0.9303(0.0202)& 0.8846(0.0060)   \\
     $5\times 5$  &  500 & 0.0292(0.0169) &0.8489(0.1417) &0.9673(0.0212)  &0.8563(0.0085) \\
          &  1000 &  0.0076(0.0042) &0.9738(0.0186) & 0.9933(0.0046)   &0.9913(0.0014)\\
     \hline
   \end{tabular}
\label{table:3}
\end{table}

Tables \ref{table:1}, \ref{table:2} and \ref{table:3} report the empirical normalized mean squared error, $F_1$-score, $d_0$ and $Acc_{all}$. Their standard errors are give in parentheses across all three colored models.
%For the star graph with $p=30$, the empirical normalized mean squared error increases a little bit as the sample size $n$ increase. That means our algorithm did not give a good performance when $p$ is large for the star graph. But, when $n$ is large enough, our algorithm have a perfect performance for selecting the best model according to the values of $F_2$-score, $d_0$ and $Acc_{all}$.
As suggested by Tables 1-3, the penalized composite likelihood method performs well in precision matrix estimation in terms of MSE over underlying star graphs, cycle graphs and grid graphs. The overall accuracy $Acc_{all}$
takes a relatively high value across all scenarios. It shows our method correctly identifies the conditional relationships and symmetry structures. The overall performance of the penalized composite likelihood is better as the sample size $n$ increases and worse as the parameter number $p$ increases.

\section{Real data analysis}
In this section, we apply our model selection method to a real glioblastoma cancer dataset. Glioblastoma multiforme (GBM) is one of the most common and aggressive forms of malignant brain cancer in adults. Despite notable  the advances of modern medicine, the overall prognosis for most GBM patients remains extremely poor. The median duration of survival is about one year \citep{ohgaki2005epidemiology}. We aim to construct a colored graphical gene regulatory network of GBM patients. Based on colored graphical Gaussian models, a gene regulatory network can be identified directly from the precision matrix \citep{werhli2006comparative}. If two genes have a direct regulatory interaction, the corresponding element of the precision matrix is non zero. The estimated colored graphical model can be used to detect genes that play pivotal roles in the development and progression of cancer. For example, the model can help to detect genes that have interactions with many other genes. Such genes are likely to play a significant role in controlling other genes' expression. In addition, the estimated model can help to identify potential mutated genes that have interactions with other genes vary significantly \citep{mohan2014node}.

\begin{figure}[htbp]
    \hspace{-4mm}
    \begin{minipage}{0.45\linewidth}
        \centering
        \includegraphics[width=3in]{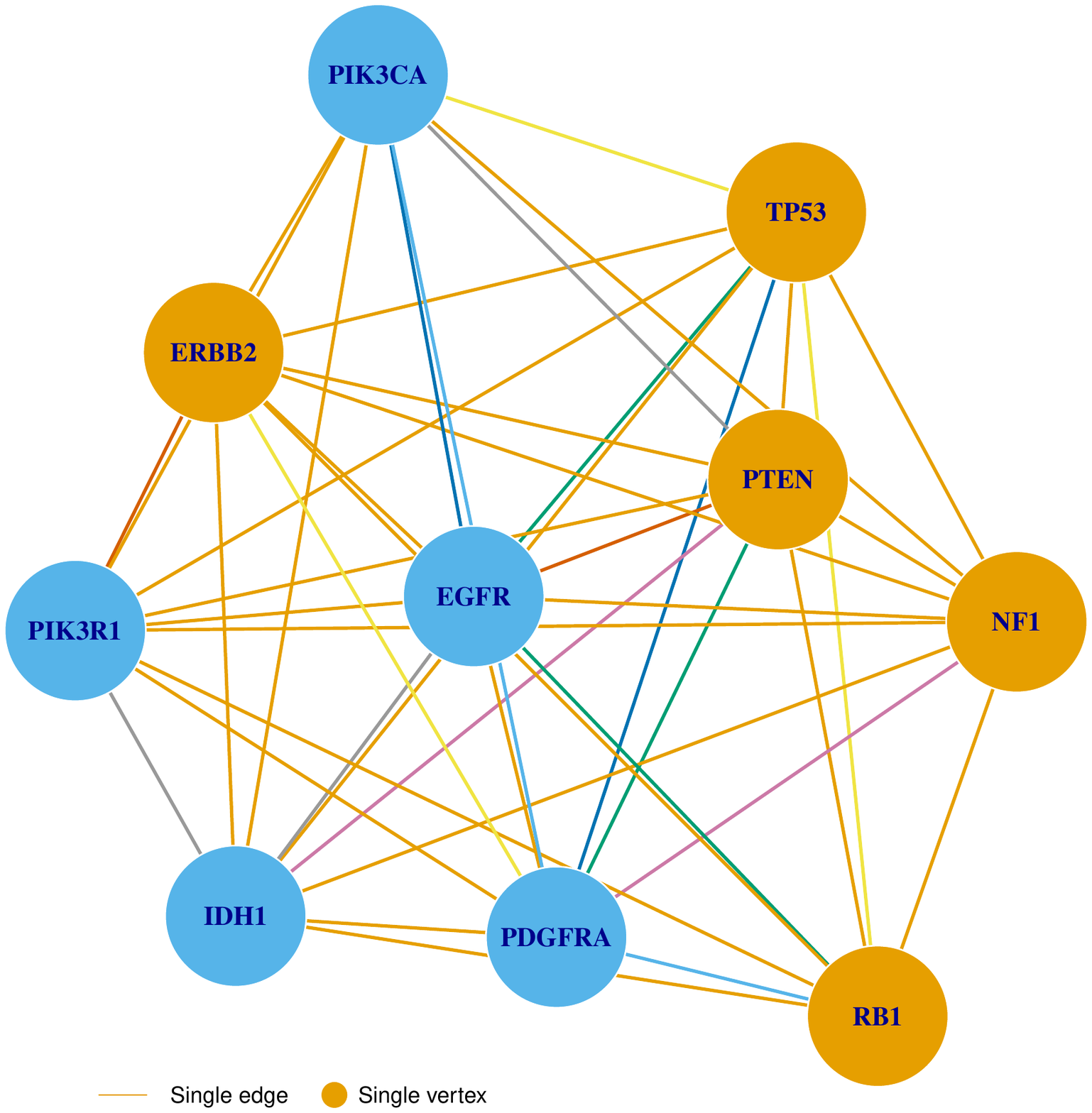}\\
        (a)
    \end{minipage}
    \begin{minipage}{0.45\linewidth}
        \centering
        \includegraphics[width=3in]{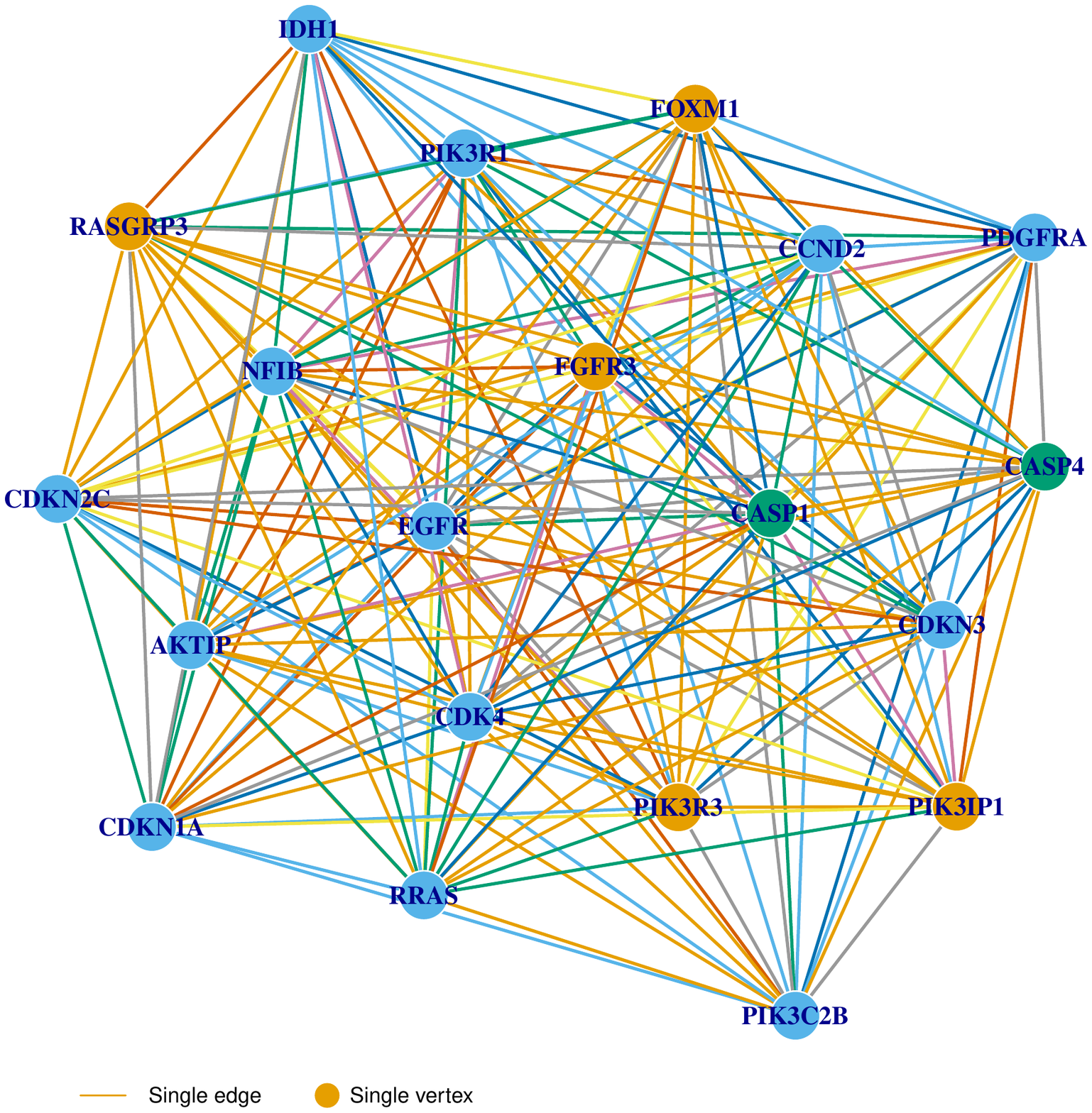}\\
        (b)
    \end{minipage}
    \caption{Estimated gene networks from our model selection method applied to the glioblastoma cancer data. In each network, color classes with a single element are displayed in gamboge. (a) 10 genes; (b) 20 genes.}
    \label{fig:2}
\end{figure}

For the purpose of the current analysis, we consider the publicly available gene expression data set downloaded from The Caner Genome Altas (TCGA) website (https://www.\\cancer.gov/about-nci/organization/ccg/research/structural-genomics/tcga). The raw \\gene expression data were generated using the Affymetrix GeneChips technology and normalized by using robust multichip average.
We first consider a gene expression dataset that consists of 200 GBM and two normal brain samples \citep{verhaak2010integrated}. We focus on 10 genes shown in Figure \ref{fig:2} (a), which have been identified to be frequently mutated in glioblastoma. Next, we consider 20 genes which are related to cell signaling pathways and play important roles in cell cycle regulation \citep{gao2016estimation}. These genes are EGFR, PDGFRA, FGFR3, RASGRP3, RRAS, PIK3C2B, PIK3R1, PIK3R3, PIK3IP1, AKTIP, NFIB, CDKN3, CDK4, CDKN1A, CDKN2C, CCND2, CASP1, CASP4, IDH1, FOXM1. The estimated colored graph for the 20 genes is shown in Figure \ref{fig:2} (b).

\begin{figure}[htbp]
    \hspace{-4mm}
    \begin{minipage}{0.9\linewidth}
        \centering
        \includegraphics[width=5in]{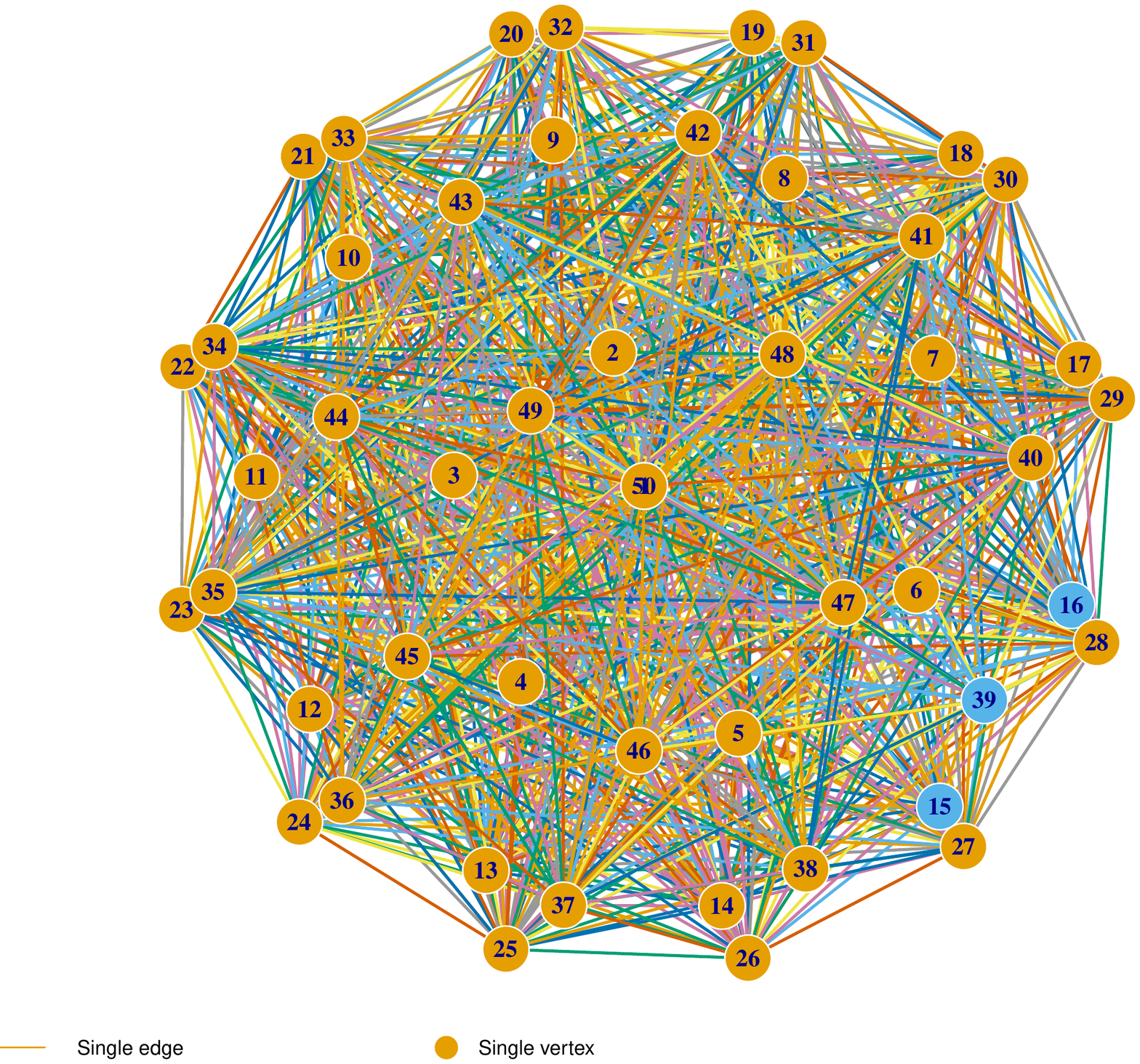}\\
    \end{minipage}
    \caption{Estimated gene networks corresponding to 50 random genes from our model selection method applied to the glioblastoma cancer data. Color classes with a single element are displayed in gamboge.}
    \label{fig:3}
\end{figure}

Finally, we consider the 840 gene signature which is reported by \cite{verhaak2010integrated}. To reduce the dimensionality of our analysis, we randomly select 50 genes from the 840 gene expressions of the 173 core samples. In our experiment, we choose the optimal set of parameters by minimizing the $BIC_{c}$ score.
In the high-dimensional colored graphical models where $p$ is extremely large, calculation of $BIC_{c}$ values over a
four dimensional grid for all $\lambda_{1}$, $\lambda_{2}$, $\lambda_{3}$ and $\tau$ may be computationally expensive. Following \citet{danaher2014joint}, we suggest a dense search over $\lambda_{1}$, $\lambda_{2}$, $\lambda_{3}$ and $\tau$. In particular, we let $\lambda_{2}$, $\lambda_{3}$ and $\tau$ be fixed at small values and conduct a line search over $\lambda_{1}$.  With tuned $\lambda_{1}$ and small values of $\lambda_{3}$ and $\tau$, we conduct a line search over $\lambda_{2}$. The dense searches for $\lambda_{3}$ and $\tau$ are the same. The estimated colored graph for the random 50 genes is shown in Figure \ref{fig:3}.

%Therefore, the estimated colored graphical model using our model selection method can help us to interpret similarities and differences of the conditional gene relationships given micro-RNAs.

%\begin{figure}[htbp]
%    \hspace{-4mm}
%    \begin{minipage}{0.9\linewidth}
%        \centering
%        \includegraphics[width=5in]{random50}\\
%    \end{minipage}
%    \caption{Estimated gene networks corresponding 50 random genes from our model selection method applying to the Glioblastoma Cancer Data. In each network, the color class with a single elements are displayed with gamboge color.}
%    \label{fig:3}
%\end{figure}

%
%Figure 6 shows the estimated conditional inverse covariance structure of genes given micro-RNAs. This structure is obtained from the model using our proposed DPS method. Black edges represent the common structure shared among all subgroups while gray edges represent unique structures to some subgroups. Verhaak et al. [7] claimed that FGFR3, PDGFA, and EGFR are all classical genes in sense that they tend to be highly expressed only in classical subtype. Thus, it is expected that they have some connectivity among them. However, in Fig. 6 from our results, none of them are connected for all subtypes. This implies that in all subtypes, they can be conditionally independent given other genes once we take out the effects of given 50 micro-RNAs on them even though they are marginally correlated.
%The sample mean is centred to be zero.

\section{Final remarks}

In this study, we propose an estimation procedure based on composite likelihood for colored graphical models. The precision matrix estimation procedure is constructed by $L_1$ penalty functions and nonconvex optimization methods. The penalized composite maximum likelihood approach offers a flexible method to estimate the underlying dependency structure and the symmetry structure. Empirical results suggest that our regularization method works efficiently for the precision matrix estimation.

The precision matrix estimation scheme based on composite likelihood includes various statistical techniques such as regularization analysis, multivariate analysis and optimization analysis. Future research topic would be extend our method to a wide variety of parameter estimation problems such as estimating parameters in hierarchical discrete loglinear models or estimating the covariance matrix in multiple undirected graphical models.

Another important topic to investigate is the asymptotic properties of the penalized regularization method. In our numerical examples in Section 4, the MSE of the estimate approaches 0 as $n$ increases. In fact, \citet{Shen12} showed that the coefficient estimate goes to 0 as $n\rightarrow \infty$ in linear regression. It is worth considering such asymptotic behaviour for colored graphical Gaussian models.

%\section{Soft thresholding operator (Notes)}
%
%We obtain the solution $\beta = b-a\frac{d|\beta|}{d\beta}$.
%
%If $\beta>0$, we have $\frac{d|\beta|}{d\beta} = 1$, $\beta=b-a>0$. Thus $b>a$.
%
%If $\beta<0$, we have $\frac{d|\beta|}{d\beta} = -1$, $\beta=b+a<0$. Thus $b<-a$.
%
%In the above two cases, we have $|b|>a$. Therefore, we also have that $sign(\beta) = sign(b)$ (Recall $a >0$).
%
%If $\beta = 0$, the subdifferential $\frac{d|\beta|}{d\beta}$ is the interval $[-1,1]$, then $0\in b-a[-1,1]$. That implies $b\in [-a,a]$, therefore, we have $|b|\leq a$.
%
%Putting all together, we get
%
%\[
% \beta =
%  \begin{cases}
%   0 & \text{if } |b| \geq a \\
%   b-a*sign(b)       & \text{if } |b| > a
%  \end{cases}\\
%  =ST(b,a) = sign(b)(|b|-a)_{+}
%\]
%
%If $|b|-a >0, ST(b,a) = sign(b)(|b|-a) = b-a$ if $b>0$ or $b+a$ if $b<0$.
%
%If $|b|-a \leq 0, ST(b,a) = 0$
%
%The central idea of these approaches is to estimate the precision matrix or the inverse of the covariance matrix which provides a conditional correlation interpretation among variables in the graph, where zero partial correlation implies pairwise conditional independence.

\bibliographystyle{apalike}
\bibliography{BayesianModelSelectionRef}

%\begin{thebibliography}{}
%
%\bibitem{Fan04}
%Shen, X., Huang, H. C., and Pan, W. (2012). Simultaneous supervised clustering and feature selection over a graph. Biometrika, 99(4), 899-914.
%
%\bibitem{Lau08}
%H{\o}jsgaard, S., and Lauritzen, S. L. (2008). Graphical Gaussian models with edge and vertex symmetries. Journal of the Royal Statistical Society: Series B (Statistical Methodology), 70(5), 1005-1027.
%
%\bibitem{Gao15}
%Gao, X., and Massam, H. (2015). Estimation of Symmetry-Constrained Gaussian Graphical Models: Application to Clustered Dense Networks. Journal of Computational and Graphical Statistics, 24(4), 909-929.
%
%\end{thebibliography}{}

\end{document}